\documentstyle[12pt]{article}
\begin{document}
\def\singlespacing{\baselineskip=16pt}
\def\doublespacing{\baselineskip=18pt}

\singlespacing
\pagestyle{empty}
\noindent
cond-mat/9605346 \hfill Published in Phys.\ Rev.\ E {\bf 50}, 2523 (1994)

\bigskip
\bigskip
\bigskip
\bigskip

\doublespacing
\begin{center}
\begin{large}
{\bf PHASE ORDERING DYNAMICS OF COSMOLOGICAL MODELS} \\
\end{large}
\bigskip
\medskip
{\em J. A. N. Filipe\footnote{Present address: BioSS, The University of
Edinburgh, James Clerk Maxwell Building, The King's Buildings, Edinburgh
EH9 3JZ, United Kingdom} and A. J. Bray} \\
\medskip
Department of Theoretical Physics \\
The University, Manchester M13 9PL, United Kingdom \\
\bigskip
\bigskip
{\bf ABSTRACT} \\
\end{center}

The ordering dynamics of the Higgs field is studied, using techniques
inspired by the study of phase ordering in condensed matter physics,
as a first step to understanding the evolution of cosmic structure
through the formation of topological defects in the early universe.
The common feature of these different physical processes is {\em scaling}.
A fully analytical approximate scheme $-$ the linear-gaussian approach
$-$ is proposed to evaluate 1-point, 2-point, etc.\ scaling functions
for the ordering dynamics of the $O(n)$-symmetric Higgs-field models.

\bigskip
\bigskip


\newpage
\pagestyle{plain}
\pagenumbering{arabic}

\newcommand{\k}{\bb{k}}
\newcommand{\r}{\bb{r}}
\newcommand{\rA}{\bb{r_1}}
\newcommand{\rB}{\bb{r_2}}

\newcommand{\at}{a(t)}
\newcommand{\aet}{a(\eta)}
\newcommand{\eA}{\eta_1}
\newcommand{\eB}{\eta_2}
\newcommand{\eZ}{\eta_0}
\newcommand{\lt}{L(t)}
\newcommand{\leA}{L(\eA)}
\newcommand{\leB}{L(\eB)}

\newcommand{\p}{\phi}
\newcommand{\pr}{\p(\r)}
\newcommand{\pre}{\p(\r,\eta)}
\newcommand{\vp}{\vec{\p}}
\newcommand{\vpr}{\vp(\r)}
\newcommand{\vprt}{\vp(\r,t)}
\newcommand{\vpre}{\vp(\r,\eta)}
\newcommand{\vpk}{\vp_{\k}}
\newcommand{\vpK}{\vp_{-\k}}

\newcommand{\mre}{m(\r,\eta)}
\newcommand{\mA}{m(1)}
\newcommand{\mB}{m(2)}
\newcommand{\vm}{\vec{m}}
\newcommand{\vmre}{\vm(\r,\eta)}
\newcommand{\vmA}{\vm(1)}
\newcommand{\vmB}{\vm(2)}
\newcommand{\M}{|\vm|}
\newcommand{\MA}{|\vmA|}
\newcommand{\MB}{|\vmB|}
\newcommand{\SA}{S_0(1)}
\newcommand{\SB}{S_0(2)}
\newcommand{\SAB}{S_0(1)S_0(2)}
\newcommand{\CZ}{C_0(1,2)}
\newcommand{\Tre}{T(\r,\eta)}

\newcommand{\cre}{C(\r,\eta)}
\newcommand{\creAB}{C(\r,\eA,\eB)}
\newcommand{\cAB}{C(1,2)}
\newcommand{\ci}{C_{\ift}}
\newcommand{\ciAB}{\ci(1,2)}
\newcommand{\cire}{\ci(\r,\eA,\eB)}
\newcommand{\cG}{C(\gm)}
\newcommand{\cg}{C_{\gm}}
\newcommand{\cgAB}{C_{\gm}(1,2)}
\newcommand{\cgg}{C_{\gm\gm}}
\newcommand{\cggg}{C_{\gm\gm\gm}}
\newcommand{\dgg}{D_{\gm\gm}}
\newcommand{\dggg}{D_{\gm\gm\gm}}
\newcommand{\fin}{f_{\ift}}

\newcommand{\ptlAvpt}{\frac{\ptl\vp}{\ptl t}}
\newcommand{\ptlAvpe}{\frac{\ptl\vp}{\ptl\eta}}
\newcommand{\ptlBvpe}{\frac{\ptl^2\vp}{\ptl\eta^2}}
\newcommand{\ptlce}{\frac{\ptl C}{\ptl\eta}}
\newcommand{\ptlcABe}{\frac{\ptl\cAB}{\ptl\eta}}

\newcommand{\intk}{\int \! \frac{d^3\k}{(2\pi)^3}}
\newcommand{\inty}{\int \! d^3\bb{y}}
\newcommand{\intinfy}{\int^{\ift}_{-\ift} \! dy}
\newcommand{\exprk}{e^{i\,\r\cdot\k}}
\newcommand{\expxy}{e^{i\,\bb{x}\cdot\bb{y}}}

\newcommand{\eq}{\equiv}
\newcommand{\al}{\alpha}
\newcommand{\dlt}{\delta}
\newcommand{\sq}{\sqrt}
\newcommand{\ift}{\infty}
\newcommand{\prl}{\parallel}
\newcommand{\ptl}{\partial}
\newcommand{\nb}{\bb{\nabla}}
\newcommand{\vdot}{\!\cdot\!}
\newcommand{\gm}{\gamma}
\newcommand{\gmre}{\gm(\r,\eta)}
\newcommand{\gmAB}{\gm(1,2)}
\newcommand{\gmi}{\gm_{\ift}}
\newcommand{\gmire}{\gmi(\r,\eA,\eB)}
\newcommand{\gmiAB}{\gmi(1,2)}
\newcommand{\FAB}{{\cal F}(1,2)}
\newcommand{\FiAB}{{\cal F}_{\ift}(1,2)}
\newcommand{\aP}{\frac{\al+1}{2}}
\newcommand{\aM}{\frac{\al-1}{2}}
\newcommand{\pp}{\!+\!}
\newcommand{\mm}{\!-\!}

\newcommand{\aloeta}{\frac{\al}{\eta}}
\newcommand{\Half}{\frac{1}{2}}
\newcommand{\bp}{\bar{p}}
\newcommand{\brho}{\bar{\rho}}
\newcommand{\aver}[1]{\left<#1\right>}
\newcommand{\curv}[1]{\left(#1\right)}
\newcommand{\brak}[1]{\left\{#1\right\}}
\newcommand{\Ju}{J_{\nu}}

\newcommand{\nn}{\nonumber}
\newcommand{\bc}{\begin{center}}
\newcommand{\ec}{\end{center}}
\newcommand{\be}{\begin{equation}}
\newcommand{\ee}{\end{equation}}
\newcommand{\ba}{\begin{eqnarray}}
\newcommand{\ea}{\end{eqnarray}}
\newcommand{\bb}[1]{{\bf #1}}
\newcommand{\rf}[1]{(\ref{#1})}
\newcommand{\bsk}{\bigskip}
\newcommand{\msk}{\medskip}
\newcommand{\nin}{\noindent}

\bsk
\nin \bb{I. INTRODUCTION} \\
When a system is rapidly quenched from a disordered phase of high
symmetry to a multi-phase region of lower symmetry it undergoes a
spontaneous symmetry breaking (SSB) phase transition.
During this transition the system develops a spatial
structure of randomly distributed domains which grow with time.
This phase ordering process has been extensively studied in the context
of condensed matter systems \cite{REV}, especially those with a
non-conserved order parameter (model A) \cite{HH}, described by the
time-dependent Ginzburg-Landau (TDGL) equation.
There is much evidence that in the late stages of growth
these systems enter a scaling regime \cite{SCALING}, in which the
two-point correlation function has the scaling form
\be
   C(\r,t_1,t_2) \eq \aver{\vp(\bb{x},t_1)\vdot\vp(\bb{x}+\r,t_2)}
                = f\left(\frac{r}{L(t_1)},\frac{L(t_2)}{L(t_1)}\right)\ ,
                           \label{C}
\ee
where $\vp$ is the vector order parameter field, $L(t)$ is the
characteristic length scale at time $t$ after the quench, $f$ is a
scaling function, and angled brackets indicate an average over initial
conditions (and thermal noise, if present).

A similar kind of phase-ordering phenomenon is believed to have occurred in
the early universe. While the big-bang theory has been widely verified by
observations (confirming that the universe began in a very hot, dense
state and has expanded and cooled down ever since \cite{W}), the origin of
cosmic structure remains unexplained.
According to the isotropy of the microwave background radiation (left
over from the early matter-radiation decoupling transition) the early
matter distribution was very uniform.
How did the universe evolve from its primordial smooth state to its
present state of lumpiness, where matter concentrates in galaxies and
galaxy clusters \cite{KT} forming a very-large scale structure?
It is believed that tiny large-scale density fluctuations, present at
decoupling, could, if strong enough, have resisted the overall expansion
and grown under gravitational collapse.
Matter in the overly dense regions of space would have clumped together
to produce general structure.
What was the origin of these small fluctuations, however, and how could
they have generated the kind of large-scale structure we see today?
Based on a process central to unified theories of particle physics $-$
that as the early universe cooled down a hypothetical field, the
{\em Higgs field}, underwent a SSB transition $-$ it has been suggested
\cite{K} that the consequent field ordering and defect formation could
have provided the mechanism to generate structure.
Field defects would form unavoidably, because `vacuum' configurations
above the horizon scale are uncorrelated. Since the defects carry
energy they could provide the fluctuations around which matter
would aggregate \cite{KT,C,ST}.

The purpose of this paper, is to use some of the techniques developed in
the framework of `model A' dynamics (i.e.\ the TDGL equation)
to study the Higgs model ordering kinetics.
This problem is technically more difficult than model A because the
equation of motion describes a damped wave propagation rather than a
purely dissipative process.
However, these non-conserved field ordering processes are likely to
exhibit similarities at late-times, where a scaling regime is expected
to occur \cite{TS,PRS,PST}. A difference, though, is that here the
characteristic length scale grows {\em linearly} with time,
$L(t)\sim t$, while $L(t)\sim\sqrt{t}$ for model A.

While domain growth phenomena, governed by the kinetics of topological
defects, have been fairly well understood within model A dynamics, a
first principles calculation of the scaling functions has proved to be a
most difficult task, and various closed-approximation schemes to evaluate
the scaling function $f(x,q)$ of equation (\ref{C}) have been proposed
in the past few years \cite{OJK,M}.
The key technique, exploited by several authors \cite{OJK}-\cite{BH},
is to to introduce a mapping $\vp(\r,t)=\vp(\vm(\r,t))$ between the
order parameter field and an auxiliary field $\vm(\r,t)$ which has, near
a defect, the physical interpretation of a position vector relative to
the defect.
With this new variable, the problem of describing the field at each
instant of time is transformed into a problem of describing the
evolution and statistics of the defect network. This approach enables
the use of a physically plausible and mathematically convenient gaussian
distribution for $\vm$. Such a distribution is unacceptable for the
order parameter itself, since this is effectively discontinuous at the
domain size scale.
The application of this sort of approach to the scalar-field model A has
recently received a critical review by Yeung et al.\ \cite{YOS}.
Methods based on a description of the wall dynamics lead to an
approximate {\em linear} equation for $m(\r,t)$, or for its correlator
$\aver{m(\bb{x},t_1)m(\bb{x}+\r,t_2)}$.
A different and promising approach, due to Mazenko \cite{M}, aims at
deriving a closed {\em non-linear} equation for $C(\r,t_1,t_2)$, built
on the equation of motion for the scalar-field model A, and the
assumption that the field $m$ is gaussian distributed at all times.
It has the virtue of yielding results with a non-trivial dependence on
the spatial dimension $d$ and it is also easily extensible to
$O(n)$-component systems.
Despite the uncontrolled nature of the gaussian assumption these
approaches have been shown to give good results, displaying most of the
expected physical features \cite{M,BH}. For the nonconserved dissipative
dynamics of model A, it has been argued that the gaussian approximation
becomes exact in the limit of large spatial dimension $d$, while for fixed
$d$ it provides the starting point for a systematic treatment \cite{BH93}.
It is also correct for any $d$ in the limit of large $n$. For the Higgs
field model considered here, the gaussian approximation is again exact
for large $n$, but the large-$d$ limit does not seem to be simple.
Nevertheless, by incorporating topological defects in a natural way,
the gaussian field approach provides the simplest non-trivial approximation
scheme for the dynamics of phase ordering.

In section 3 we attempt to apply the Mazenko approach to the $O(n)$-field
Higgs model. The late-time pair correlation function is then given by the
Bray-Puri-Toyoki (BPT) function \rf{BPT} \cite{BPT}, as a function
$C(\gm,n)$ of the normalized correlator of $\vm$, $\gm(\r,t_1,t_2)$,
which obeys an approximate non-linear equation. The BPT function embodies
the asymptotic defect structure, while $\gm(\r,t_1,t_2)$ describes the
dynamical dependence of $C(\r,t_1,t_2)$.
The mapping used by Mazenko \cite{M,BH}, however, restricts the field to
evolve within the bound $|\vp|\leq 1$, which is incompatible with the
oscillatory bulk relaxation of the Higgs field, and leads to an
inconsistent approach. The difficulties with this approach, however,
motivate our next attempt to tackle the problem.
In section 4 we consistently eliminate the field bulk oscillations
by restricting the asymptotic field dynamics to the `vacuum manifold'.
Extending Mazenko's gaussian approach to the Non-Linear Sigma Model
(NLSM), using the unit vector mapping $\vp=\vm/\M$, the pair correlation
function is still given by the BPT function, but now $\gm$ obeys a
different approximate equation.
Rather then solving numerically this equation for $\gm$, which is rather
complicated, in section 5 we propose a fully analytical approximate scheme
to evaluate $C$. This amounts to replacing $\gm$ by its large-$n$ limit
in the argument of the BPT function, but keeping the remaining
$n$-dependence unchanged.
As $n\to\ift$ the equation for $\gm$ becomes linear and exactly
solvable \cite{TS}, so we may call this scheme the Linear-Gaussian
(LG) approach.
Although we cannot use the NLSM for a scalar field, the approach still
holds for this case if one takes $n=1$ in the BPT function.
In section 6 the LG approach will be generalized to evaluate other kinds
of scaling functions, such as the average of the energy density and its
correlation function.

\bsk
\nin \bb{II. THE HIGGS FIELD MODEL} \\
In this section we briefly review basic notions about the cosmological
background model. The Higgs field model is presented and its dynamics
are discussed.

As is usual practice, we shall consider a {\em flat} expanding universe
as the model for the bulk cosmological background \cite{W,KT}.
In this case the local curvature is zero and the metric is
space-independent, given by
\be
 ds^2 = c^2\,dt^2 - \at^2 d\r^2 = \aet^2 \curv{c^2\,d\eta^2 - d\r^2} \ ,
                        \label{metric}
\ee
where: $t$ and $\r$ are {\em comoving} coordinates (i.e. the reference
frame is moving with the cosmic flow);
$\at$, or $\aet$, is the space {\em expansion factor}$\,$;
$\eta$, the {\em conformal time}, defined by
\be
                    d\eta \eq dt/\at \ ,
                        \label{eta}
\ee
plays the role of `real time' in a static universe:
the {\em horizon} of an `event' after a time $t$ $-$ its maximum
range of influence after time $t$ $-$ is given by
$h(t)=\int^h_0 dr=\int^t_0 c\:dt/\at = c\int^\eta_0 d\eta = c\:\eta(t)$.
In a flat universe, the function
\be
  \al(\eta) \eq 2\: \frac{d\ln a(\eta)}{d\ln \eta} \ ,
                        \label{alpha}
\ee
varies slowly with time from $\al = 2$ (radiation era) to $\al = 4$
(matter era). Away from the matter-radiation decoupling transition $\al$
can be regarded as a constant and the expansion factor is given by a
power-law $a \sim t^{\al/(\al+2)}$ $\sim \eta^{\al/2}$ \cite{KT,TS}.

In the early {\em radiation dominated era} the energy was dominated by
relativistic particles (with equation of state $\bp=\brho/3$), yielding
$a \sim t^{1/2}$ $\sim \eta$ and $\brho \sim a^{-4}$ $\sim t^{-2}$.
Here $\bp$ and $\brho$ are the uniform background pressure and energy
density.
Once the universe cooled down and matter decoupled from radiation this
became the dominant source of gravitation (with negligible pressure
$\bp\ll\brho$), yielding
$a \sim t^{2/3}$ $\sim \eta^2$ and $\brho \sim a^{-3}$ $\sim t^{-2}$
in the {\em matter dominated era}.
As matter became transparent to radiation, the matter perturbations
started to grow.

A simple class of SSB theories is provided by the (real) $n$-component
Higgs field models, where a `global' $O(n)$ symmetry is broken
\cite{KT,TS,PST}.
These theories include several cases where topological defects form:
domain walls ($n=1$), global strings ($n=2$), global monopoles ($n=3$)
and global textures ($n=4$), which are of potential interest as a
mechanism to generate cosmic structure.

The dynamics of the Higgs field $\vprt \eq (\p^1,...,\p^n)$ in an
expanding universe is derived from the Lagrangian density
\cite{KT}
\be
  {\cal L}(\r,t) =
 \Half\:\curv{\ptl\vp/\ptl t}^2 -\Half\:\curv{\nb\vp/\at}^2 - V(\vp) \ ,
                        \label{Lagrangian}
\ee
where $\nb$ is with respect to comoving coordinates, and $V(\vp)$ is a
generalized 'double-well' potential with an $O(n)$ symmetric `vacuum
manifold' where $\vp^2=1$.
Minimizing the action $S=\int dt \int d^3\r\, \at^3\, {\cal L}(\r,t)$
(where $dt\: d^3\r\, \at^3$ is the covariant 4-volume element)
with respect to variations of $\vp$, and using conformal time, yields
the equation of motion
\be
  \ptlBvpe + \aloeta\;\ptlAvpe =
                        \nb^2\vp - \aet^2\:\frac{\ptl V}{\ptl\vp} \ ,
                        \label{phi.eq}
\ee
a wave equation with a damped `friction force'
$(\al/\eta)(\ptl\vp/\ptl\eta)$, which mimics expansion in the comoving
frame (and destroys the Lorentz invariance), and a non-linear force
$\ptl V/\ptl\vp$ which drives the field to the `vacuum manifold'.
The initial conditions $-$ corresponding to a disordered state before
the SSB phase transition $-$ shall be discussed in the appendix, where
we present the solution of \rf{phi.eq} in the limit $n\to\ift$.
The price for using conformal time is to have an effective potential
in equation \rf{phi.eq} with a time-dependent amplitude.
The particular form of the potential, however, should not affect in any
essential way the late-time dynamics and scaling properties.
We expect, for instance, the main effect of $\aet^2$ to be a decrease
by a factor $1/a$ in the comoving size of the defects core, which simply
speeds up the system entry into the scaling regime.
To simplify the subsequent discussion we shall from now on discard the
$a^2$ factor in the equation of motion \rf{phi.eq}.
We will not really need that for computational purposes, as we
shall be using the NLSM.

Taking conformal time on the same footing as real time, equation
\rf{phi.eq} can be viewed as a `general-relativistic analogue' of the
TDGL equation, describing the dynamics of non-conserved systems.
The Higgs Hamiltonian density corresponding to \rf{Lagrangian}
\be
  {\cal H}(\r,t) = \frac{1}{\at^2}\,\brak{\,
  \Half\,\curv{\ptl\vp/\ptl\eta}^2+\Half\,\curv{\nb\vp}^2 + V(\vp)} \ ,
                        \label{Hamiltonian}
\ee
is (apart from $1/a^2$) similar to that of a static (Minkowski)
universe, and compared to the TDGL model has an extra `centripetal'
term $(\ptl\vp/\ptl\eta)^2$.
For a vector field in the `vacuum manifold' it leads to an energy
density which decays (due to expansion and dissipation) like the
background, $\brho \sim 1/t^2$.
Therefore, the Higgs field yields density fluctuations of constant
amplitude $\delta\rho/\brho = (\rho-\brho)/\brho \sim {\rm const}$
which, through Einstein's equations, provide a {\em source} for
perturbations in the matter distribution.

Assuming the existence of a late-time {\em scaling} regime (which
has been confirmed by numerical simulations \cite{TS,PRS,PST}), the
dimensional analysis of \rf{phi.eq} leads to a {\em characteristic
scale} growing with the horizon
\be
                L(\eta) \ \sim\ \ c\:\eta \ ,
                        \label{L}
\ee
implying that the field defects move with relativistic speed.
We therefore expect the pair correlation function \rf{C} to take
the asymptotic scaling form $C(\r,\eA,\eB)=f(x,q)$, with scaling
variables $x=r/\eA$ and $q=\eB/\eA$, where $r=|\rA-\rB|$ is the distance
between the two points.

Causality constrains the field correlations after the SSB transition.
Two field configurations at times $\eA$ and $\eB$ can only be causally
correlated if their distance $r$ is below the sum of their horizons
$c\,\eA$ and $c\,\eB$ (i.e.\ if the horizons intersect).
Therefore, the condition for $C(\r,\eA,\eB)\neq 0$ is (taking $c=1$)
\be
                       r \ < \ \eA + \eB \ .
                        \label{causality}
\ee
If one of the horizons contains the other configuration
($\eA$ or $\eB > r$) the correlations are `direct'.
Otherwise, `indirect' correlations can occur through common causal
correlations with intermediate points in the region of intersection of
the horizons.

Unlike purely relaxing systems, the wave nature of the Higgs dynamics
forces the late-time saturating field not to satisfy $|\vp|<1$ even if
its initial condition does.
To see how the field tends to its 'vacuum manifold' we linearize
equation \rf{phi.eq} as $\vp$ approaches a given `vacuum' state $\vp_0$.
Considering a single-domain region where $\vp$ can be taken as
uniform, and noticing that the only restoring force is parallel to
$\vp_0$ (normal to the manifold) due to the symmetry of the `vacuum', we
find, at late-times
\be
    \vp(\eta)\cdot\vp_0 \ \simeq \ 1 - \frac{1}{\aet}\:
    \brak{c_1(\al)\,\cos(A\eta) \ + \ c_2(\al)\,\sin(A\eta)} \ ,
                        \label{oscillations}
\ee
where $A^2=\curv{\ptl^2V/\ptl(\vp^2)^2}_1$,
and $c_1(\al)$ and $c_2(\al)$ are arbitrary constants.
For a scalar field $\vp(\eta)\cdot\vp_0$ is replaced by $|\p(\eta)|$.
We conclude that the Higgs field saturates with damped oscillations.

\bsk
\nin \bb{III. GAUSSIAN THEORY FOR A `SOFT' FIELD} \\
In this section we apply to the Higgs model the gaussian approach
proposed by Mazenko \cite{M} for the TDGL dynamics.
Although the approach, which is based on an unphysical mapping for the
Higgs dynamics, leads to an inconsistent theory, it will motivate the
implementation of a gaussian approach for a unit vector field in section 4.

To derive an equation for the pair correlation function \rf{C}, we
multiply the equation of motion \rf{phi.eq}, evaluated at point
$(1)\eq(\rA,\eA)$, by $\vp(2)\eq\vp(\rB,\eB)$ and average over the
ensemble of initial conditions, yielding the exact equation
\be
  C(\ddot{1},2) + \frac{\al}{\eA}\:C(\dot{1},2) = \nb^2\:\cAB + \FAB \ ,
                        \label{exact.C.eq}
\ee
where the driving force, or non-linear (NL) term, is
\be
    \FAB = - \aver{\vp(2)\vdot\frac{\ptl V}{\ptl\vp(1)}} \ ,
                        \label{NL}
\ee
and $C(\dot{1},2)\eq\ptl\cAB/\ptl\eA=\aver{\dot{\vp}(1)\vdot\vp(2)}$, etc.
To transform \rf{exact.C.eq} into a closed equation we need to write
the NL term as some approximate non-linear function of $\cAB$.
A key idea, exploited by several authors within the TDGL dynamics
\cite{OJK,M,BH}, is to employ a non-linear mapping between the order
parameter $\vpre$ and an auxiliary `smooth' field $\vmre$.
This new variable describes the late-time defect network structure,
which will have formed at the late stages of field ordering, and
allows for the approximation to be implemented.
The most obvious way to define the function $\vp(\vm)$, is to follow
Mazenko's suggestion \cite{M} of using the {\em equilibrium profile}
equation of an isolated defect (in a comoving frame)
\be
               \nb^2_{m}\,\vp = \ptl V/\ptl\vp \ ,
                        \label{soft.map}
\ee
with boundary conditions $\vp(0)=0$ and $\vp(\vm)\,\to\,\vm/|\vm|$ as
$(|\vm|\to\ift)$, and where $\nb_m$ is the gradient with respect to $\vm$.
Close enough to a defect (i.e. for $|\vm|\ll L(\eta)$, where the
field is unaffected by neighboring defects) $\vmre$ can be identified
as the comoving position vector of point $\r$ from the (nearest part of
the) defect. This picture requires, of course, that $n\leq d$.
With \rf{soft.map} the magnitude of $\vp(\vm)$ is a monotonically
increasing function of the magnitude $|\vm|$, approaching for large
$|\vec{m}|$ the `attractor' value $1$ imposed by the potential.
For a scalar field, the function $\phi(m)$  has a typical sigmoid form.

The mapping \rf{soft.map} restricts the field magnitude to be $|\vp|<1$.
This is appropriate for diffusion fields evolving from a disordered
state, but is physically incorrect for the Higgs field dynamics, where
the system self-organizes oscillating about the `vacuum' states, as
shown by \rf{oscillations}.
While we can prove that the use of \rf{soft.map} leads to an
inconsistent theory \cite{U1}, it seems unlikely that an adequate
one-to-one mapping could be defined for this problem.
In section 4 we shall overcome this technical difficulty by
restricting the field dynamics to the `vacuum manifold', $\vec{\phi}^2=1$.
Meanwhile, for completeness we will pursue this approach a little further
using \rf{soft.map} to derive a closed equation for $\cAB$.

Following Mazenko \cite{M}, we now assume that $\vmre$ is a gaussian
random field (with zero mean) at all times, described by the pair
distribution function
\be
\!\! P(\vmA,\vmB) = N^n \:\exp \brak{ - \frac{1}{2(1-\gm^2)}
      \curv{ \frac{\vmA^2}{\SA} + \frac{\vmB^2}{\SB} -
             \frac{2\gm\,\vmA\vdot\vmB}{\sq{\SAB}} } \! }
                        \label{P}
\ee
\ba
    \SA = \aver{\mA^2} & , & \CZ = \aver{\mA\,\mB} \ ,
                        \label{moments} \\
                \gmAB  & = & \frac{\CZ}{\sq{\SAB}} \ ,
                        \label{gm}
\ea
where $N=\curv{2\pi\sq{(1-\gm^2)\SAB}}^{-1}$, and $m(1)$ and $m(2)$
are the same (arbitrary) component of $\vmA$ and $\vmB$.
All the averages over the ensemble of initial $\vp$ configurations are
replaced by gaussian averages on $\vm$, and can be evaluated as functions
of the second moments $\SA$, $\SB$, and $\gmAB$.
However, from \rf{L} and the mapping \rf{soft.map}, according to which
$\vm$ can be identified as a position vector, we anticipate the asymptotic
scaling form
\be
    S_0(\eta) = \frac{2\,\eta^2}{\beta} \ \sim \ L(\eta)^2 \ ,
                        \label{S0.scale}
\ee
where $\beta$ is a parameter to be determined.
Thus, within a gaussian approach, the only variable in the problem is the
function $\gmAB$, which accounts for the particular dynamics of $\vp$.

The driving force \rf{NL} in equation \rf{exact.C.eq} is then given, as
a non-linear function of $\cAB$, by Mazenko's result \cite{M,BH}
\be
      \FAB \ = \ - \ 2\,\frac{\ptl\,C(\gm)}{\ptl\,\SA}
           \ = \ \frac{\beta}{2}\,\frac{\gm\,\cg(\gm)}{\eA^2} \ ,
                        \label{gauss.NL}
\ee
where $\cg\eq(\ptl C/\ptl\gm)$ and we have used \rf{S0.scale}.
Note that, by use of the mapping (\ref{soft.map}), there is no longer
any explicit dependence on the potential $V(\vec{\phi})$ in
(\ref{gauss.NL}), though the relation between $\vec{\phi}$ and $\vec{m}$
depends on $V$.
At late-times the field will be saturated almost everywhere except at
the defect cores (whose size is much smaller than the domain scale), and
we may, for simplicity, evaluate the gaussian averages replacing the
profile mapping \rf{soft.map} by its discontinuous asymptotic form
$\vp(\vm)=\vm/\M$, for a vector field, or $\p(m)={\rm sign}\,(m)$,
for a scalar field. At late times, therefore, the detailed form of
the potential is not important (although it must, of course, have the
`Mexican hat' form in order to support non-trivial solutions of
(\ref{soft.map})). This is in accord with the expected `universal' nature
of the late-stage scaling behavior.

Evaluating the pair correlation function $\cAB$, using \rf{P} and
the mapping above, yields the explicit relation $C=C(\gm,n)$,
which we will call the `BPT function' \cite{BPT},
\ba
  C(\gm)
  & = &\aver{ \frac{\vmA}{\MA}\vdot\frac{\vmB}{\MB} }_m
                        \label{CG} \\
  & = & \gmAB\:
       \frac{n\curv{B\curv{\frac{n+1}{2},\frac{1}{2}}}^2}{2\pi}\,
       F\curv{ \frac{1}{2},\frac{1}{2};\frac{n+2}{2};\gmAB^2 }\ ,
                        \label{BPT}
\ea
where $B(x,y)$ is the beta function and $F(a,b;c;z)$ is the
hypergeometric function.
The substitution of \rf{gauss.NL} and $C(\gm,n)$ into equation
\rf{exact.C.eq} yields the approximate closed equation for $\gmAB$,
which for a vector field must be regarded as the independent variable.
In the limit $n\to\ift$ the BPT function reduces to $C(\gm,\ift)=\gm$,
yielding $\FAB=\cAB/\SA$, and \rf{exact.C.eq} becomes a linear equation.

We now focus on the pair correlation function at {\em equal-times}
($\eA=\eB=\eta$), which is of interest by itself and also yields the
initial condition to solve the general equation \cite{BH}.
Equation \rf{exact.C.eq} then reads
\be
  \Half\:\ddot{C}(1,2) - C(\dot{1},\dot{2}) + \Half\aloeta\:\dot{C}(1,2)
  = \nb^2\cAB + \FAB \ ,
                        \label{equal.CD.eq}
\ee
where $\dot{C}(1,2)\eq\ptl C/\ptl\eta$, etc.
The unknown quantity $C(\dot{1},\dot{2})$ may be eliminated to get
a third order equation in $C$.
Then, replacing $\FAB$ by its approximate form \rf{gauss.NL}, using
\rf{S0.scale}, and looking for an isotropic scaling solution
$\cre=f(x)$, which implies $\gm(\r,\eta)=\gm(x)$, with $x=r/\eA$, leads
to the equation for $\gm(x)$:
\ba
\! \! \! \! \! \! & \! \! \! & \! \! \! \!
\frac{x(4-x^2)}{2}\brak{ \gm'''+ 3\gm'\gm''\dgg + \gm'^3\dggg } +
\curv{x^2\,\frac{3(\al-2)}{2}+2(d+1-\al)}\!\brak{\gm''+\gm'^2 \dgg}\nn \\
\! \! \! \! \! \! & \! \! \! & \! \! \! \!
- \curv{ \frac{x}{2}(\al-2)(2\al-3) + \frac{2}{x}(\al-1)(d-1) } \,\gm'
\ = \ - \beta\,\brak{ (1-\al)\,\gm + x(1+\gm \dgg)\,\gm' }
                        \label{equal.gm.eq}
\ea
where $\gm'=d\gm/dx$, etc, $\dgg(\gm)=\cgg/\cg$, $\dggg(\gm)=\cggg/\cg$,
and $\cg\eq\ptl C/\ptl\gm$, etc.
The NL functions $\dgg(\gm)$ and $\dggg(\gm)$ are obtained from \rf{BPT}
and embody all the $n$-dependence of \rf{equal.gm.eq}.
The boundary conditions for equation \rf{equal.gm.eq} are
$\gm(0)=1$, from definition \rf{moments}, $\gm'(0)=0$, from
$\gm(x)=1-O(x^2)$ as $x\to 0$, and $\gm(2)=0$, from $C(\gm)\sim\gm$ as
$\gm\to 0$ and the causal condition $f(x)=0$ for $x\geq 2$.
We notice that the boundary points are both singular, which makes
the numerical solution of \rf{equal.gm.eq} difficult.

For a {\em scalar field} the BPT function \rf{BPT} can be inverted to
give $\gm=\sin\curv{\pi\,C/2}$, yielding a NL term
$\FAB=(2/\pi\SA)\,\tan\curv{\pi\cAB/2}$. Hence we can express
\rf{equal.gm.eq} as an explicit non-linear equation for the scaling
function $\cre=f(x)$:
\ba
   &   & \frac{x(4-x^2)}{2}\,f(x)'''
    + \curv{x^2\frac{3(\al-2)}{2}+2(d+1-\al)} \,f(x)'' \nn \\
   &   & - \ \curv{\frac{x}{2}(\al-2)(2\al-3)
                 + \frac{2}{x}(\al-1)(d-1)} \,f(x)' \nn \\
   & = & - \ \beta\brak{
            \frac{2}{\pi}\,(1-\al)\,\tan\curv{\frac{\pi}{2}f(x)}
          + x\,\sec^2\curv{\frac{\pi}{2}f(x)} \,f'} \ .
                        \label{equal.f.eq}
\ea
To perform a small-$x$ expansion of \rf{equal.f.eq}, we recall that
with the mapping $\p={\rm sign}(m)$, used to evaluate $C(\gm)$, the
condition $f(0)=\aver{\p^2}=1$ has been built into the theory (although
in an inconsistent manner).
We find that $f(x)$ admits a series in odd powers of $x$ (implying
that all derivatives at $x=0$ are determined without recursion), giving
the linear behaviour, or Porod's regime \cite{P},
\be
   f(x) = 1 - \frac{1}{\pi}\,\sq{\frac{2\beta\al}{(\al-1)(d-1)}}\:x
          + O(x^3) \ \ \ ,  \ \ \ (x\to 0) \ ,
                        \label{small-x.f}
\ee
which is a physical consequence of having `sharp' walls at late times.
To find the small-$(2-x)$ asymptotic form of $f(x)$, we notice that as
$f(x)\to 0$ equation \rf{equal.f.eq} becomes linear and has three
independent solutions.
Since the singularity at $x=2$ is regular we try a Frobenius power
series solution \cite{A},
$A_0(2-x)^p(1+\sum_{k=1}^{\ift}a_k(2-x)^k)$, and find that the
equation admits a leading power-decay $f(x)\sim (2-x)^{p}$ as $x\to 2$,
where $p$ can assume any of the values: $p=0$, $p=1$ or $p=\al+(d-1)/2$.
$p=0$ must be excluded as being incompatible with the boundary
conditions (it would imply $A_0=0$), and thus the solution has the
general asymptotic form as $x\to 2$:
\be
    f(x) \sim A^{(1)}_0\,(2-x)^{\al+(d-1)/2}\brak{1+O(2-x)}
      + A^{(2)}_0\,(2-x)\brak{1+O(2-x)}  \ .
                        \label{general.decay}
\ee
Since the BPT function \rf{BPT} has the same behaviour $C(\gm,n)\sim\gm$
(and $\dgg$, $\dggg\to 0$) as $\gm\to 0$ or $n\to\ift$, to linear order
in the regime $x\to 2$ and $f\sim\gm$ equation \rf{equal.gm.eq} is
$n$-independent and identical to its large-$n$ limit.
In the appendix we discuss the large-$n$ limit of the NLSM \rf{NLSM}
and find that $\fin(x)\sim (2-x)^{\al+1}$ as $x\to 2$, for $d=3$.
Therefore, for any value of $n$, and at least for short-ranged
initial conditions, we expect the leading power-law decay
\be
   f(x) \ \sim \ (2-x)^{\al+\frac{d-1}{2}} \ \ \ , \ \ \ (x\to 2) \ .
                        \label{f.decay}
\ee
Although we are not looking to solve equation \rf{equal.f.eq}, we
describe how one in principle could do it.
{}From \rf{small-x.f} and \rf{causality}, the boundary conditions are:
$f(0)=1$, $f'(0)=-(1/\pi)\sq{2\beta\al/(\al-1)(d-1)}$ and $f(2)=0$.
The parameter $\beta$ is numerically determined by imposing the
coefficient of the dominant solution in \rf{general.decay} to vanish,
$A^{(2)}_0(\beta)=0$.
In the large-$n$ limit, where equation \rf{exact.C.eq} becomes linear
and the gaussian approach is exact, $\beta$ can be found analytically.
Comparing \rf{exact.C.eq} with the linear equation \rf{gmi.eq}, which
amounts to compare the limit of \rf{gauss.NL},
$\FiAB=\beta_{\ift}\gmAB/2\eA^2$, with $\aver{T(1)}$ given by
\rf{Ti.scale}, yields
\be
              \beta_{\ift} = 2\,T_0 = 3\,(2\al+1)/2 \ .
                        \label{betai}
\ee
In conclusion, although the mapping \rf{soft.map} discards the field
oscillations \rf{oscillations} and leads to an inconsistent theory
\cite{U1}, equation \rf{equal.gm.eq}-\rf{equal.f.eq}, despite its
intrinsic incorrectness bears no obvious signs of inconsistency.
A Porod's regime \rf{small-x.f} is obtained as a consequence of the
'sharp' wall constraint $\p(m)={\rm sign}(m)$ used to evaluate
$C(\gm,n)$.
We have shown that the manner in which $f(x)$ vanishes at $x=2$, given by
\rf{f.decay}, is independent of $n$ and exact.

\bsk
\nin \bb{IV. GAUSSIAN THEORY FOR THE NON-LINEAR SIGMA MODEL} \\
In this section we study the dynamics of a vector Higgs field within
the NLSM. By constraining the field to lie on the vacuum manifold,
this model automatically avoids the technical difficulties associated
with the asymptotic bulk oscillations noted in section 3.
We develop a gaussian approach, analogous to that of section 3, and
derive an approximate equation for $\cAB$.

Long after the SSB phase transition the driving potential $V$ closely
confines the Higgs field to the `vacuum manifold' almost everywhere
(except at the field defect cores).
We have shown, however, that the wave nature of the dynamics leads to a
field bulk saturation accompanied by slow decaying oscillations about
the 'vacuum state', preventing us to define an adequate one-to-one
mapping between $\vp$ and an auxiliary field $\vm$.
The mapping \rf{soft.map}, for instances, forces the field to obey
$|\vp|\leq 1$ at all times and yields an inconsistent approach.
To overcome this technical problem, we notice that the oscillations
\rf{oscillations} are unlikely to have a major effect on the late-time
dynamics of the field defect network (and thus on the scaling
properties), and may thus be consistently discarded by restricting the
$O(n)$ field dynamics to the vacuum manifold.
Replacing the vanishing driving force $\ptl V/\ptl\vp$ in \rf{phi.eq}
by a non-linear coupling term which constrains the length of the field,
the field evolution is now described by the non-linear sigma model
(NLSM) equation \cite{TS}:
\be
  \ptlBvpe + \aloeta\;\ptlAvpe = \nb^2\vp + \Tre\,\vp \ ,
                        \label{NLSM}
\ee
where $\Tre$ is the free Lagrangian density in \rf{Lagrangian}
\be
          \Tre \eq \curv{\ptlBvpe -\nb^2\vp}\!\vdot\vp
                =  \curv{\nb\vp}^2 - \curv{\ptlAvpe}^2 \ .
                        \label{T}
\ee
As another advantage of using the NLSM, the ordering dynamics becomes
independent of the details of the potential $V(\vp)$ and, in particular,
the factor $\aet^2$ in \rf{phi.eq} is suppressed.

The exact equation for the pair correlation function $\cAB$ is still
given by \rf{exact.C.eq}
\be
  C(\ddot{1},2) + \frac{\al}{\eA}\:C(\dot{1},2) = \nb^2\:\cAB + \FAB \ ,
                        \label{exact.C.eqh}
\ee
where, from \rf{NLSM} and \rf{T}, the NL term is now given by
\be
    \FAB = \aver{T(1)\,\vp(1)\vdot\vp(2)} \ ,
                        \label{NLh}
\ee
which must be replaced by some approximate non-linear function of
$\cAB$ in order to transform \rf{exact.C.eq} into a closed equation.
Following the strategy of section 3, we introduce a non-linear mapping
between the order parameter $\vpre$, which is now not well defined near
the defects, and an auxiliary `smooth' field $\vmre$.
We can no longer define $\vp=\vp(\vm)$ using the equilibrium profile
equation of an isolated defect \cite{M}, which yields a trivial relation
everywhere except at the defect cores where it is singular.
The natural way to define the relation between the unit vector $\vpre$
and $\vmre$ amounts to replacing \rf{soft.map} by its discontinuous
asymptotic form \cite{TurokNote}
\be
                  \vp(\vm) = \frac{\vm}{|\vm|} \ .
                        \label{hard.map}
\ee
This mapping only determines $\vmre$ up to a factor (which e.g., may be
a function of time), and there is now no obvious physical interpretation
for the new variable. Up to a factor, however, we may still regard
$\vmre$ as a position vector (close enough to a defect) like in section 3.

For mathematical convenience we assume that $\vmre$ is a gaussian random
field (with zero mean) at all times, described by the pair distribution
function \rf{P}-\rf{moments}.
All the averages over the ensemble of initial $\vp$ configurations are
replaced by gaussian averages on $\vm$, and can be evaluated as
functions of $n$, $\SA$, $\SB$, and $\gmre$, the normalized
$m$-correlator, which contains all the dynamic dependence.

In the same spirit which lead to expression \rf{gauss.NL} in section 3,
using the mapping \rf{hard.map} and the gaussian property of $\vm$ we can
shown \cite{U2} that the NL term \rf{NLh} is then given, as an
approximate non-linear function of $\cAB$, by
\ba
    \FAB & = & \brak{ \aver{\dot{m}(1)^2} - \aver{(\nb\mA)^2} }\:
               2\,\frac{\ptl C}{\ptl\SA}
      + \dot{S_0}(1)^2\:\frac{1}{3}\,\frac{\ptl^2 C}{\ptl\SA^2} \nn \\
         & + & \brak{ (C_0(\dot{1},2))^2 - (\nb\CZ)^2 }\:
               \frac{1}{3}\,\frac{\ptl^2 C}{\ptl\CZ^2} \nn \\
         & + & \dot{S}_0(1)C_0(\dot{1},2)\:\frac{1}{3}\,
               \brak{ \frac{2n}{n-1}\,\frac{\ptl^2 C}
                                 {\ptl\SA\ptl\CZ}+Q_n(1,2) }
                        \label{gauss.NLh}
\ea
\ba
Q_n(1,2) & = & (n-3)(n-1)\aver{\curv{\MA^3\MB}^{-1}}_m
                        \label{Q} \\
         & = & \frac{1}{\SA\sq{\SAB}}\,
  \frac{(n-1)\curv{ B \curv{\frac{n-1}{2},\frac{1}{2}}}^2}{2\pi}\,
  F\curv{\frac{1}{2},\frac{3}{2};\frac{n}{2};\gmAB^2 } \ .
                        \label{gauss.Q}
\ea
Using \rf{moments}, \rf{gm} and \rf{BPT} the NL term \rf{gauss.NLh} can
be fully expressed in terms of $\gm$ and $S_0$.
For example, the correlators
$\aver{\dot{m}(1)^2}=$ $C_0(\dot{1},\dot{2})_{2\to 1}$ and
$\aver{(\nb\mA)^2}=$ $-\nb^2 C_0(1,2)_{2\to 1}$, the derivative
$\ptl C/\ptl\SA=-\cg\,\gm/2\SA$, where $\cg(\gm)\eq\ptl\cg/\ptl\gm$,
and similarly for the other derivatives of $\cAB$.
Substituting the NL term and the BPT function \rf{BPT} into
\rf{exact.C.eqh} we get the equation for $\gmAB$, which is the
independent variable. Specializing to equal-times ($\eA=\eB=\eta$), and
looking for an isotropic scaling solution $\gm(\r,\eta)=\gm(x)$, we then
obtain an approximate closed equation for $\gm(x)$, the NLSM version of
\rf{equal.gm.eq}, the boundary conditions for which have been given in
section 3.
Even if we take $S_0$ to be time independent, this equation will still
be much more complicated than \rf{equal.gm.eq}.

If $\vm$ in the NLSM is set to have the same interpretation as in section
3, and thus to obey \rf{S0.scale}, we may compare the NL terms
\rf{gauss.NLh} and \rf{gauss.NL}. The NLSM gaussian approach generates
the `soft' field result, as long as $\aver{(\nb m)^2}=1$, plus
additional terms following from the consistent use of the mapping
\rf{hard.map}.
This differences indicate that the gaussian approach is not quantitatively
accurate, since \rf{phi.eq} and \rf{NLSM} should yield equivalent
asymptotic dynamics.

\bsk
\nin \bb{V. LINEAR-GAUSSIAN APPROXIMATION} \\
Rather then solving the extremely complicated approximate non-linear
equation for $\gmAB$, we propose a fully analytical scheme $-$ the
`Linear-Gaussian' (LG) approach $-$ to evaluate $\cAB$, which combines
a gaussian mapping for a unit vector $\vp$ with the large-$n$ exact
solution.

We notice that the relation $C=C(\gm,n)$, defined by \rf{CG} and given
by the BPT function \rf{BPT} for a gaussian $\vm$, accounts effectively
for the presence of the field defects (through the orientation of
$\vp=\vm/\M$), and also for their topological nature (through the
$n$-dependence), and so it already describes fairly well the late-time
defect structure.
Hence, the particular form of the function $\gmAB$, which contains the
dynamical dependence of $\cAB$, should not be so relevant and may be
approximated rather crudely. For simplicity, we replace
$\gm$ by $\gmi$, the exact solution in the large-$n$ limit \cite{KYGNote}.
The scaling function $f_{LG}(x_s,q)=\cAB_{LG}$
with $n=1,..,4,\ift$ and $\al=2$ and 4,
obtained using this procedure, is plotted in figures 1 and 2
with fixed values of $q=\eta_2/\eta_1$ and abscissa $x_s=2r/(\eta_1+\eta_2)$,
and in figures 3 and 4 with fixed values of $x_s$ and abscissa $q$.
More details are given in section 7 and in the figure captions.

As $n\to\ift$, $\M=\curv{\sum_i m_i^2}^{1/2}\to\sqrt{n\,S_0}$, and it is
easy to find the limit of the functions $Q_n$ and $\cG$, either from
their definitions \rf{Q} and \rf{CG} or from their gaussian averages
\rf{gauss.Q} and \rf{BPT}. The BPT function reduces to
$C(\gm,n)\to\gmi=\ci$ and $Q_n(1,2)\to 1/\SA\sq{\SAB}$.
Equations \rf{exact.C.eqh}-\rf{NLh} yield the self-consistent linear
equation
\be
      \gmi(\ddot{1},2) + \frac{\al}{\eA}\:\gmi(\dot{1},2)
      = \nb^2\:\gmiAB + \aver{T(1)}\:\gmiAB \ ,
                        \label{gmi.eq}
\ee
\be
                   \aver{T(1)} = T_0/\eA^2 \ ,
                        \label{Ti.scale}
\ee
where the scaling form \rf{Ti.scale} follows from a dimensional analysis
of \rf{gmi.eq} or \rf{T} (and from translational invariance), and the
constant $T_0$ is to be found self-consistently (see \rf{T0} in the
appendix).
The linear term $\FiAB=\aver{T(\eA)}\,\gmiAB$ is the limit of the
previous NL term:
the gaussian expression \rf{gauss.NLh}, or the definition \rf{NLh} and
\rf{T}, where $\vp\to\vm/\sq{n\,S_0}$.

Instead of determining $\gmiAB$ by solving the linear equation
\rf{gmi.eq} at equal times, which (like \rf{equal.CD.eq}) is third
order, it is easier to calculate the correlation function of the exact
large-$n$ solution of the NLSM equation \rf{NLSM}, which is second order.
Equation \rf{gmi.eq} for $\gmi=\ci$ can be derived from the large-$n$
limit of \rf{NLSM} (just like \rf{exact.C.eqh} was derived from
\rf{NLSM}), so the two procedures to obtain $\gmi$ are equivalent.
Turok and Spergel \cite{TS} have solved the large-$n$ NLSM in momentum
space and determined the structure factor corresponding to a random
initial field.
In the appendix we present and Fourier-transform their result to
3-dimensional real space, yielding the scaling function
$\fin(x,q)\eq\cire=\gmire$:
\be
\fin(x,q) = \frac{\theta(1+q-x)}{N}\:
                \frac{1}{x\,q^{\al+1/2}}\:
                \int^{B}_{A}ds\:s(1-s^2)^{\aM}(q^2-(x-s)^2)^{\aP} \ ,
                        \label{fi}
\ee
where
\be
           x = r/\eta_1 \ \ \ , \ \ \ q = \eta_2/\eta_1 \ \ \ ,
                        \label{xq}
\ee
\ba
    (B,A) & = & (x+q,x-q)   \ \ , \ \ x \leq 1-q \nn \\
          & = & (1,  x-q)   \ \ , \ \ |1-q| \leq x \leq 1+q \nn \\
          & = & (1,   -1)   \ \ , \ \ x \leq q-1 \ ,
                        \label{AB}
\ea
and $N=4/5,\;32/63$ for $\al=2,\;4$. At equal-times,
\be
\fin(x,1) = \frac{\theta(2-x)}{N}\:
                \frac{1}{x}\: \int^{1}_{x-1}ds\:
        s(1-s^2)^{\frac{\al-1}{2}}(1-(x-s)^2)^{\frac{\al+1}{2}} \ .
                        \label{equal.fi}
\ee
The small-$x$ behaviour of $\fin(x,1)$ can either be obtained from the
large-$n$ limit of the gaussian equation for $\gm$, e.g.\
\rf{equal.gm.eq}, or by expanding \rf{equal.fi} as $x\to 0$.
Both procedures yield the leading behaviour as $x\to 0$
\ba
\fin(x,1) = & 1 - (5/8)\:\ln\curv{1/x}\:x^2 + \cdots & \ \ \al=2 \nn \\
          = & 1 - (27/16)\:x^2 + \cdots  & \ \ \al=4 \ .
                        \label{small-x.fi}
\ea
Expanding the BPT function \rf{BPT} as $\gm\to 1$ and using
\rf{small-x.fi} yields the small-$x$ expansion for the equal-times pair
correlation function within the LG approach.
For a scalar field we have
\ba
f_{LG}(x,1) = & 1 -
\frac{1}{\pi}\sq{5\ln\curv{1/x}}\:x + \cdots & \ \ \al=2,\ \ n=1 \nn \\
            = & 1 -
\frac{1}{\pi}\sq{27/2}          \:x + \cdots & \ \ \al=4,\ \ n=1 \ ,
                        \label{small-x.fLG.one}
\ea
and for a vector field
\ba
f_{LG}(x,1) = & 1 - A_1(x)\:x^2 + \cdots & \ \ \al=2,\ \ n>1 \nn \\
            = & 1 - A_2\:x^2 + \cdots & \ \ \al=4,\ \ n>1 \ ,
                        \label{small-x.fLG.n}
\ea
where $A_1(x)=(5/8)(\ln(1/x))^2$ and $A_2=(27/16)\ln(1/x)$ for $n=2$, and
$A_1(x)=(5/4)\ln(1/x)$ and $A_2=27/8$ for $n=3$.
Performing a small-$(1+q-x)$ expansion of \rf{fi} we find the
leading power-law decay
\be
f(x,q) \simeq \fin(x,q) \simeq
\frac{B\curv{\aP,\Half}}{4(\al+1)\,B\curv{\al,3/2}}\:
\frac{\curv{\frac{q+1}{2}}^{\al+1}}{q^{\al/2}}\:(1+q-x)^{1+\al} \ ,
\ \ \ (x\to q+1) \ .
                        \label{large-x.f}
\ee
In the limit of very-different times ($\eA\ll\eB$), we obtain the
leading time-decay
\be
f(x,q) \simeq \fin(x,q) \simeq
\frac{1}{q^{3/2}}\:\frac{B\curv{\aP,3/2}}{B\curv{\al,3/2}} +
O(1/q^{7/2}) \ , \ \ \ (q\to\ift) \ .
                        \label{large-q.f}
\ee
By the same arguments discussed in section 3, the asymptotic forms
\rf{large-q.f} and \rf{large-x.f} (the different-times generalization of
\rf{f.decay}) are exact and the same for all $n$.
In fact, as $x\to 1+q$ or $q\to\ift$ and $\gm\to 0$ equation
\rf{exact.C.eqh} becomes the linear equation \rf{gmi.eq}, from which
the same powers, but not the amplitudes, can be obtained.

\bsk
\nin \bb{VI. OTHER SCALING FUNCTIONS IN THE LG APPROACH} \\
The LG method, implemented in section 5 to evaluate the pair correlation
function, can be extended to other scaling functions.
In this section we evaluate the pressure, the average energy
density and the energy density correlation function.

As long as we replace $\vp$ by its saturation form $\vm/\M$ (or $\p$ by
$m/|m|$, for a scalar field), the scaling functions will have built in
the late-time defect structure. Treating $\vm$ as a gaussian field, the
dynamical dependence of the scaling functions is again embodied by
$\gm(\r,\eA,\eB)$.
In the same spirit as in section 5, we replace $\gm$ by its large-$n$ limit.
In short, we keep in the $n$-dependence of the scaling properties
through the gaussian averages over the $\vm$ vectors,
and treat the gaussian moments of $\vm$ in the large-$n$ limit.
As mentioned in section 4, the mapping \rf{hard.map} only determines $\vm$ up
to a factor (which may be time-dependent), and thus there is some
freedom to fix the form of the second moment $S_0\eq\aver{m^2}$.
Although the choice $S_0={\rm const.}$ would greatly simplify the algebra
(e.g.\ reducing the number of pair contractions of gaussian averages
containing $\dot{\vp}$), we find it physically more convenient to regard
$|\vm|$ as a length (close to a defect), and thus to keep the scaling
form \rf{S0.scale}, i.e. $S_0=2\,\eta^2/\beta$. When written in terms of
$\gm$, though, the results are independent of the choice made.

The Higgs field {\em energy density} (see \rf{Hamiltonian}) and
isotropically averaged {\em pressure} are given by \cite{KT}
\ba
\rho(\eta) & = & \frac{1}{2\,\aet^2}\,\aver{
       \dot{\vp}^2 + \curv{\nb\vp}^2 } + \aver{V(\vp)}
                        \label{energy} \\
p(\eta)    & = & \frac{1}{2\,\aet^2}\,\aver{
       \dot{\vp}^2 - \frac{1}{3}\,\curv{\nb\vp}^2 } - \aver{V(\vp)} \ ,
                        \label{pressure}
\ea
and scale as $1/t^2$, like the background $\bar{\rho}$ and $\bar{p}$.
To evaluate $\rho$ and $p$ within the LG approach, we first consider a
vector field. In this case, the potential term is negligible (and
identically zero in the NLSM) and can be ignored.
Writing the derivatives of $\vp$ in terms of the derivatives of $\vm$,
expanding the gaussian average into a sum of pair contractions,
expressing the averages containing $\nb_m\vp$ as the limit $2\to 1$ of
derivatives of $\cAB$ with respect to the gaussian moments,
and treating $\vm$ in the limit $n\to\ift$, yields
\ba
 & \aver{\dot{\vp}^2} = \cg(1,1)\,\curv{\frac{\aver{\dot{m}^2}}{S_0} -
       \curv{\frac{\dot{S_0}}{2\,S_0}}^2}
        = \cg(1,1)\,\aver{\dot{\vp}^2}_{\ift}
                        \label{dtdt} \\
 & \aver{\curv{\nb\vp}^2} = \cg(1,1)\,\frac{\aver{(\nb m)^2}}{S_0}
        = \cg(1,1)\,\aver{\curv{\nb\vp}^2}_{\ift} \ ,
                        \label{nbnb}
\ea
where $\cg(1,1)\eq(\ptl\cAB/\ptl\gmAB)_{2\to 1}=(n-1)/(n-2)$ for $n\geq
3$, $\cg(1,1)=\ln(L/w)$ to leading order for $n=2$ (where $w$ is the
string size, introduced as a short-distance cut-off), and
\ba
 \aver{\dot{\vp}^2}_{\ift} & = & \gmi(\dot{1},\dot{2})_{2\to 1}
   \eq \gmi(\dot{1},\dot{1}) = \frac{T_0}{(\al-2)\eta_1^2} \ , \nn \\
 \aver{(\nb\vp)^2}_{\ift} & = & (\nb_1\nb_2\,\gmi(1,2))_{2\to 1}
   \eq -\nb^2\gmi(1,1) = (\al-1)\aver{\dot{\vp}^2}_{\ift} \ ,
\ea
where we used \rf{limits} and $\ci=\gmi$.
Hence, from \rf{energy}-\rf{nbnb}, the LG approach gives
\be
\rho(\eta) = \frac{n-1}{n-2}\:\rho_{\ift}(\eta) \ \ \ , \ \ \
   p(\eta) = \frac{n-1}{n-2}\:p_{\ift}(\eta) \ \ \ , \ \ \ n>2 \ ,
                        \label{rho.rhoi.n}
\ee
\be
\rho(\eta) = \ln(L/w)\:\rho_{\ift}(\eta) \ \ \ , \ \ \
   p(\eta) = \ln(L/w)\:p_{\ift}(\eta) \ \ \ , \ \ \ n=2 \ ,
                        \label{rho.rhoi.two}
\ee
with
\be
\rho_{\ift}(\eta)=\frac{\al\,T_0}{2(\al-2)}\,\frac{1}{a^2\eta^2}
\ \ \ , \ \ \
p_{\ift}(\eta)=\frac{(4-\al)\,T_0}{6(\al-2)}\,\frac{1}{a^2\eta^2} \ .
                       \label{rhoi}
\ee
In the radiation dominated era, where $\al=2$, $\gmi(\dot{1},\dot{1})$
and $-\nb^2\gmi(1,1)$ (and $\rho_{\ift}$ and $p_{\ift}$) have a leading
order logarithmic divergence.
Their difference, though, is finite and gives $\aver{T}=T_0/\eta^2$
(see \rf{limits}) (and also $p_{\ift}/\rho_{\ift}=1/3$).
The relevant case, however, is the matter dominated era ($\al=4$), when
matter perturbations started to grow, yielding $p(\eta)=0$ with $n\geq 2$
and
\be
\rho(\eta) = \curv{6.75 \ + \
         \frac{6.75}{n-2}}\,\frac{1}{a^2\eta^2} \ \ \ , \ \ \ n>2 \ .
                        \label{LG.rho.n}
\ee

Although a consistent implementation of the LG method requires the use
of the NLSM, and thus a vector field, the approach can be extrapolated
for a scalar field in an elegant manner.
This was already done in section 5, where we simply extended the results
for the scalar field correlation function taking $n=1$ in the BPT function
(see \rf{small-x.fLG.one}), rather than deriving an equation for
$C(1,2)$.
The difference for a scalar field is that the {\em wall width}, $w$,
plays a role in the dynamics, making the scaling functions (which
contain time-dependent prefactors) differ from their dimensional analysis
form. Moreover, the potential term in \rf{energy}-\rf{pressure} has now
a relevant contribution $\aver{V}=\aver{\curv{\nb\vp}^2}/2a^2$.
A convenient definition for the non-comoving wall width (which is
constant in time for sharp domain walls) is $w\eq 4/\sigma$, where
$\sigma$ is the non-comoving (or physical) surface tension given by
\be
   \aet\:\sigma = \int_{-\ift}^{\ift}dx\,(d\phi_w/dx)^2 \ ,
                        \label{sigma}
\ee
where here $\phi_w(x)$ represents a single planar domain wall, and $x$ is
a comoving coordinate normal to the wall. The value of $\sigma$ depends,
through $\phi_w(x)$, on the explicit form of the driving potential.
In the spirit of the LG approach we exploit the asymptotic mapping
$\p(m)={\rm sign}(m)$ to perform the gaussian averages, and treat the
$m$-correlators in the large-$n$ limit.
To evaluate $p$ and $\rho$, we write the derivatives of $\p$ in terms
of the derivatives of $m$ and expand the gaussian average into a sum
of pair contractions. Noting that $d\p/dm\eq\p'$ is sharply peaked at
$m=0$ and that $|\nb m|_{m=0}=1$, we get
$\aver{\p'^2}=\!\int_{-\ift}^{\ift}dm P(m) \p'^2= a \sigma P(0)$,
where $P(m)\!=\!e^{-m^2/2S_0}/\sqrt{2\pi\,S_0}$ is the one-point
probability distribution for $m$.
Using $\p'^2=a\sigma \delta(m)$ (which follows from \rf{sigma}) and
integrating by parts, we get
$\aver{(\p'^2)''}=a\sigma\aver{\delta''(m)}=a\sigma\,P''(0)$.
Therefore,
$\aver{\dot{\p}^2}=a\sigma\sqrt{S_0/2\pi}\,\gm(\dot{1},\dot{1})$ and
$\aver{(\nb \p)^2}=a\sigma\sqrt{S_0/2\pi}\,(-\nb^2\gm(1,1))$, which
are the analogues of \rf{dtdt}-\rf{nbnb}.
Since $\sigma$ embodies the extra physical feature of the scalar field,
we can treat the remaining factors in the large-$n$ limit. Taking
$S_0=\eta^2/T_0$ (which follows from \rf{S0.scale} and \rf{betai}),
$\gm(\dot{1},\dot{1})\to\aver{\dot{\vp}^2}_{\ift}$ and
$-\nb^2\gm(1,1)\to\aver{(\nb\vp)^2}_{\ift}$, which are then given by
\rf{limits}, we get, from \rf{pressure}-\rf{energy},
\be
\rho(\eta) = \frac{\sigma}{\sqrt{2\pi\,T_0}}\,\frac{2\al-1}{\al}\,
a\eta\,\rho_{\ift}(\eta) \ \ , \ \
   p(\eta) = \frac{\sigma}{\sqrt{2\pi\,T_0}}\,\frac{7-4\al}{3\al}\,
a\eta\,\rho_{\ift}(\eta) \ \ , \ \ n=1 \ .
                        \label{rho.rhoi}
\ee
In the radiation dominated era $\rho$ and $p$ have again a leading
logarithmic divergence. In the matter dominated era, we obtain
\be
\rho(\eta) =
\frac{\sigma}{\sq{2\pi}}\,\frac{7\sq{6.75}}{4}\,\frac{1}{a\eta} \ \ , \ \
p(\eta) = -
\frac{\sigma}{\sq{2\pi}}\,\frac{3\sq{6.75}}{4}\,\frac{1}{a\eta}
\ \ \ , \ \ \ n=1 \ .
                        \label{LG.rho}
\ee

We now look to evaluate the correlations between the energy
density terms of the Higgs field, i.e. $\dot{\vp}^2$ and $(\nb\vp)^2$.
For simplicity we shall restrict to the case of a scalar field.
Writing the derivatives of $\p$ in terms of the derivatives of $m$,
expanding the gaussian average into a sum of pair contractions,
replacing $\p'^2$ by $a\sigma\delta(m)$, doing some gaussian integrals
by parts, using \rf{P}-\rf{moments}, and treating the $m$-correlators in
the large-$n$ limit, we obtain
(with $\aver{X\,Y}_c=\aver{X\,Y}-\aver{X}\aver{Y}$)
\ba
\aver{\dot{\p}(1)^2\dot{\p}(2)^2}_c
+ A\:\gm_{11}\gm_{22} = \frac{A}{(1-\gm^2)^{5/2}} & & \nn \\
\brak{\!\Gamma\,(\gm_{11}\!-\!\gm_1^2)(\gm_{22}\!-\!\gm_2^2)
 + \gm^2\!\curv{(\gm_1\gm_2)^2\!-\!\Gamma\,\gm_{11}\gm_{22}}
 + 2\curv{\Gamma\,\gm_{12}\!+\!\gm\gm_1\gm_2}^2 } \ , & &
                        \label{stst}
\ea
\ba
\aver{(\nb\p(1))^2(\nb\p(2))^2}_c
+ A\:\Delta_1\gm\Delta_2\gm = \frac{A}{(1-\gm^2)^{5/2}} & & \nn \\
\brak{\!\Gamma\,(\Delta_1\gm\!-\!\gm_r^2)(\Delta_2\gm\!-\!\gm_r^2)
 + \gm^2\!\curv{\gm_r^4\!-\!\Gamma\,\Delta_1\gm\Delta_2\gm}
 + 2\curv{\Gamma\,\gm_{rr}\!+\!\gm\gm_r^2}^2
 + \frac{2\,\Gamma\,\gm_r^2}{r^2} } \ , & &
                        \label{srsr}
\ea
where $A\eq\sigma^2\,a(\eta_1)\eta_1a(\eta_2)\eta_2/(2\pi\,T_0)$,
$\Gamma\eq(1-\gm^2)$, and
$\gm=\gm_{\ift}(1,2)$,
$\gm_1=\gm_{\ift}(\dot{1},2)$,
$\gm_{12}=\gm_{\ift}(\dot{1},\dot{2})$,
$\gm_r=\ptl\gm_{\ift}/\ptl r$,
$\gm_{rr}=\ptl^2\gm_{\ift}/\ptl r^2$, and
$\gm_{11}=\gm_{\ift}(\dot{1},\dot{1})$,
$\Delta_1\gm=-\nb^2\gm_{\ift}(1,1)$, are given by
\rf{DelB} and \rf{limits}.
We have checked that, as expected, the results are independent of
whether we take $S_0$ to be constant or given by \rf{S0.scale}.
The scaling functions corresponding to \rf{stst} and \rf{srsr},
normalized in the form $\aver{X\,Y}_c/\aver{X}\aver{Y}$, have been plotted
in figures 5 and 6, respectively, for the matter era.
Details and comments are given in the next section and in the figure
captions.

\bsk
\nin \bb{VII. SUMMARY AND DISCUSSION} \\
Two distinct gaussian approaches for the $O(n)$ Higgs field dynamics,
in a flat expanding universe, were proposed to evaluate the pair
correlation function, and other scaling functions.
Both theories are based on a non-linear mapping between the order
parameter $\vpre$ and an auxiliary field $\vmre$, which varies smoothly
in the vicinity of the field defects.
For simplicity and mathematical convenience, $\vmre$ is assumed to be a
gaussian random field, yielding an approximate closed scheme to evaluate
the scaling functions. The field $\vp$ itself, which is effectively
discontinuous near the defects, is not suitable to be treated as
gaussian.

In the `soft' field theory of section 3, based on the equation of
motion \rf{phi.eq}, we have followed Mazenko's gaussian approach
\cite{M,BH} for model A dynamics, where the mapping is defined by the
equilibrium profile equation $\nb^2_{m}\,\vp=\ptl V/\ptl\vp$. In this
case, $\vmre$ is identified as a position vector relative to the nearest
field defect.
The mapping \rf{soft.map}, however, is incompatible with the late-time
field oscillations in the bulk \rf{oscillations} (which are absent in
purely relaxational systems). By studying the linear dynamics of the
gaussian moment $\CZ$, given by equation \rf{moments}, we can prove that
this theory is inconsistent, and therefore we have not looked to solve
numerically the rather complicated equations \rf{equal.gm.eq} or
\rf{equal.f.eq}.
The fact that, despite this intrinsic inconsistency, the pair
correlation function displays correct physical features, such as
\rf{small-x.f} and \rf{f.decay}, is not, however, a merit of the
approximation used. The small-$x$ Porod's regime for a scalar field
follows from the use of the BPT function \rf{BPT}, which has built in
the late-time defect structure, and the asymptotic power-decay, which
occurs in the linear (small-$f(x)$) regime, is universal for all $O(n)$
'soft' and 'hard' field models.

In section 4, we have developed a more consistent theory, based on the
NLSM, or 'hard' field, dynamics \rf{NLSM}.
We do not expect the field bulk oscillations \rf{oscillations} to have a
relevant effect on the scaling properties, so we consistently fix the
field magnitude to eliminate the previous mapping incompatibility.
Also, since the field now evolves on the vacuum manifold, the dynamics
are independent of the driving potential.
The auxiliary field is now defined by $\vp(\vm)=\vm/\M$. Although it can
still have the same interpretation as in section 3, $\vm$ is only
determined up to a factor and we are free to choose $<\!\vm^2\!>$.
The relation $C(\gm,n)$, between the pair correlation function and
the normalized $\vm$-correlator, is given by the BPT function \rf{BPT}
for a gaussian $\vm$, and $\gm$ obeys an approximate equation,
derivable from \rf{exact.C.eqh}-\rf{gauss.NLh}.

Rather then solving this complicated equation for $\gm$, in section 5 we
propose a fully analytical scheme to evaluate $\cAB$.
Recognizing that the BPT function captures the essential late-time
defect structure, we approximate the asymptotic field dynamics even
further replacing $\gm$ by $\gmi$, the exact solution for the limit
$n\to\ift$.
The pair correlation function is then given (in a symbolic notation) by
$\cAB_{LG}=BPT(\gmiAB,n)$.
Although the NLSM only holds for vector fields, the LG approach can
be extended to $n=1$, since it only depends on the large-$n$ dynamics,
which is the same for both equations \rf{phi.eq} and \rf{NLSM}.
In this case the scaling properties are evaluated using the mapping
$\p(m)={\rm sign}(m)$ and the gaussian assumption. The pair correlation
function, for instance, is again given by the BPT function, with $n=1$,
and by the same argument we replace $\gm$ by $\gmi$.
The scaling form $f_{LG}(x,q)$, for
$n=1,..,4,\ift$ and in the radiation and matter dominated eras, is
plotted in:
figures 1 and 2, respectively, with different fixed values of $q$ and
abscissa $x_s\eq 2r/(\eta_1+\eta_2)=2x/(1+q)$, and in
figures 3 and 4, respectively, with different fixed values of $x_s$ and
abscissa $q$. The normalization is as follows:
$f(0,1)=1$ for $q=1$; for $q\neq 1$ we used the time-dependent condition
$f(x,q)/\gmi(x,q)\to 1$ as $x\to 0$, such that curves with different $n$
cut the origin at the same point.

The LG approach for the Higgs model is the analogue of the
Ohta-Jasnow-Kawasaki approximate scheme in model A dynamics \cite{OJK}.
In that case, $\fin(x,1)=\exp(-x^2/8)$. The greater complexity
of \rf{fi}-\rf{equal.fi} is due to the causal condition \rf{causality}
which these obey.
The main physical features are preserved in this approach: the
threshold power-law behaviour \rf{large-x.f} (imposed by causality) is
exact, and for $n=1$ a linear Porod's regime, \rf{small-x.fLG.one}, is
obtained for $\al=4$.
For $\al=2$, however, we obtain a logarithmic modified Porod's regime,
$f(x,1)=1+O(x\,\ln x)$ (slightly apparent in figure 1), which is probably
an artifact of the LG approach and has no physical meaning.
This logarithmic correction is absent in the small-$x$ expansion
\rf{small-x.f} of section 3.

We have seen in section 5 that all the exact and the gaussian expressions
have the same limit as $n\to\ift$. In fact, the gaussian approach
becomes exact (for random gaussian initial conditions) in this limit
since the equation for $\vm$ becomes linear. This equation is derived
from the linearised equation for $\vp$ (replacing $\vp$ by
$\vm/\sqrt{n\,S_0}$), and its form depends on the choice made for $S_0$.
Also, the two gaussian approaches, for the `soft' field and for the
NLSM, become equivalent (and exact) as $n\to\ift$ and the LG approach
could be implemented equally well using either.
We find, however, that the NLSM provides a more systematic and
self-consistent framework for this purpose, while the `soft' field
model yields the physical motivation to employ the NLSM (and proves
useful in the LG calculation of other scaling functions with $n=1$).

In section 6 we have extended the LG approach to evaluate other scaling
properties of the Higgs field.
For these cases we do not know how to built closed approximate equations
like those of sections 3 and 4, and the method proves especially useful.
If we restricted ourselves to the gaussian approach we could express other
scaling functions in terms of $\gm$ and its derivatives, but we could
not solve for these derivatives numerically.
For a scalar field we can still use the asymptotic mapping $\p(m)=m/|m|$,
but we have to account for the non-trivial role of the wall width,
$w$, which is inversely proportional to $\sigma$, the surface tension
\rf{sigma}.
The LG results \rf{rho.rhoi.n}, \rf{rho.rhoi.two}, \rf{rho.rhoi}, give
the average field energy density \rf{energy} and pressure \rf{pressure}
as being proportional to $\rho_{\ift}$, the energy density in the limit
$n\to\ift$.
The factor of proportionality is $n$-dependent, and
is also time-dependent for $n=2$ and $n=1$.
Since $\rho_{\ift}$ and $p_{\ift}$, given by \rf{rhoi}, have a leading
logarithmic divergence at $\al=2$, we have discarded the radiation
dominated era, which is a less relevant case in the formation of cosmic
structure, and next summarize the LG results in the matter era ($\al=4$).
With $n>2$, \rf{LG.rho.n} gives $\rho=6.75(1+1/(n-2))/a^2\eta^2$,
which compared with the fit to simulation results \cite{PST}:
$6.75(1 + (20/9)/(n-2))/a^2\eta^2$
shows a fair agreement up to a factor $\sim 2$ in the the correction term.
With $n=1$, \rf{LG.rho} gives $\rho=\sigma\,{\rm const}/a\eta$, yielding
energy density fluctuations growing linearly with time $t$, rather than
having a constant value as in the vector case.
This well known result \cite{KT} means that walls, if present, would
rapidly dominate the energy of the universe.
With $n\geq 2$ we obtain zero pressure, as expected.
With $n=1$ we get a negative pressure $p=-3\rho/7$, yielding a source term
$\rho+3p<0$, which can be regarded as indicating
an effective domain wall repulsion \cite{KT} in the scaling regime,
and is a reflection of domain growth.
We recall that for an isolated equilibrium domain-wall perpendicular to
the $x$ direction (for which $\dot{\p}^2=0=\ptl\p/\ptl y=\ptl\p/\ptl z$),
the field pressure components along each axis are $p_x=0$ and
$p_y=p_z=-\rho$ \cite{KT}.
The pole of \rf{rho.rhoi.n} at $n=2$ is built in the approach through the
use of a unit vector (i.e.\ the defect core-size $w\to 0$) since,
in fact, the `sharpness' of the string cores leads to a logarithmic cut-off
given in \rf{rho.rhoi.two} \cite{Toyoki}.

Finally, we have done a LG calculation of the correlations between the
energy terms $\dot{\vp}^2$ and $(\nb\vp)^2$, which are the sources for
the perturbations in the cosmological matter distribution.
For simplicity we have restricted ourselves to the case of a scalar field.
In contrast to the vector case, $\dot{\p}^2$ cannot be regarded as the
`centripetal' energy due to the field wandering around in the `vacuum
manifold'. At late-times, though, $\dot{\p}^2$, or $(\nb\p)^2$, vanish
everywhere in the bulk regions and thus probe the presence of domain
walls (where energy is concentrated).
Using \rf{fi} and \rf{limits}, we have computed the scaling
function $\aver{\dot{\p}(1)^2\dot{\p}(2)^2}_c/
\aver{\dot{\p}(1)^2}\aver{\dot{\p}(2)^2}$, given by \rf{stst},
in the matter era.
Figure 5 shows the results with different fixed values of $q=\eta_2/\eta_1$
and abscissa $x_s\eq 2x/(1+q))$).
Remarkably, as $x$ increases from zero there is a dramatic change from
large positive values to negative values.
We interpret these set of plots as giving evidence of domain walls
dynamics (in a statistical sense):
the correlation peak (for fixed $q$) is displaced along the `distance'
axis as the time separation between the two points increases (i.e. as $q$
departs from 1).
Its amplitude decrease is dictated by statistical incoherence as the points
move apart, and its displacement, $x_{s,peak}$, must be proportional to
the typical distance traveled by a wall during the time $|\eta_1-\eta_2|$.
The equal-time ($q=1$) divergence of the peak at the origin (i.e.\ of
$\aver{\dot{\p}^4}$) is an artifact of the absence of a short-distance
cut-off in $\gmi(x,1)$ as $r$ drops below the wall width $w$.
The scaling function $\aver{(\nb\p(1))^2(\nb\p(2))^2}_c/
\aver{(\nb\p(1))^2}\aver{(\nb\p(2))^2}$, given by \rf{srsr}, is plotted
in figure 6 for the matter era. In this case the peak remains at
the origin, while its amplitude decreases, as $q$ departs from 1.
Since both energy density terms probe the presence of domain walls,
it is not very clear to us why this correlation function is so different
from the previous one shown in figure 5.
It seems that its form for $x<0.5$ is entirely dictated by its singular
prefactor $1/(1-\gmi^2)^{5/2}$, which is plotted in figure 7.

We conclude by discussing some directions for future work.
By linearizing the full equation of motion \rf{exact.C.eq},
with \rf{gauss.NL} or \rf{gauss.NLh}, for the correlation function around
the scaling solution, it should be possible to show that the scaling
solution is a stable attractor of the dynamics.
In particular, the prescaling regime (e.g.\ corrections to scaling)
can be described using the gaussian closure schemes of sections 3 or 4.
The early-time behaviour, however, is not accessible within the
auxiliary field methods utilized here, which assume a well-defined
defect structure.
The dependence on the initial state of the system before the phase
transition is also of interest.
In the present work, short-range spatial correlations in the initial
state are considered, appropriate for a disordered system in equilibrium
at high temperature.
In the context of model A dynamics, it has been shown that the asymptotic
scaling behavior is modified if sufficiently long-ranged power-law
correlations are present in the initial state \cite{BHN}.
Such correlations can also be incorporated in the LG approach, through
a modification of the large-$n$ solution presented in the appendix.
Finally, the extension of the functions \rf{stst} and \rf{srsr}
for vector fields, which involves extensive calculations, can be used to
evaluate the correlations between the matter distribution perturbations
induced by the Higgs field \cite{PST} and may, therefore, have a direct
cosmological interest.

\bsk
\nin \bb{ACKNOWLEDGEMENTS} \\
We thank N.\ Turok for many helpful discussions and suggestions,
and M.\ Birse for comments on the manuscript.
J.F.\ thanks JNICT (Portugal) for a research studentship.

\bsk
\nin \bb{APPENDIX: the large-$n$ solution of the NLSM} \\
For simplicity we shall still take $\vp^2=1$, which differs from the
usual normalization, $\vp^2=n$, used to solve large-$n$ models.

To leading order as $n\to\ift$ the NL factor \rf{T} is replaced by its
average (over initial conditions), $\Tre \to \aver{\Tre}$, which one
expects to have the scaling form \rf{Ti.scale}, i.e.
$\aver{\Tre} = T_0/\eta^2$, where $T_0$ is a constant to be determined
self-consistently.
The equation of motion \rf{NLSM} becomes linear and has been solved in
momentum space, with the following initial conditions, at some early
time $\eZ>0$ after the SSB transition \cite{TS}:
$\vp(\r,\eZ)$ is a (gaussian) random unit vector in each initial
correlation volume, i.e. its Fourier components are white noise
correlated, $\aver{\vpk(\eZ)\vdot\vpK(\eZ)}=\Delta$;
$\ptl\vpk(\eZ)/\ptl\eta\,\to\,0$ as $\k\to 0$, to ensure that
$\ptl\vp(\r,\eZ)/\ptl\eta$ respects the assumption of homogeneity of
the early universe on scales above the horizon.
The solution obtained is \cite{TS}
\ba
    \vpk(\eta) & = & A_{\nu}\:\curv{\frac{\eta}{\eZ}}^{3/2}\:
           \frac{\Ju(k\eta)}{(k\eta)^{\nu}}\:\vpk(\eZ) \ ,
                        \label{Turok.phi} \\
               \nu & = & 1 + \al/2
                        \label{nu} \\
               T_0 & = & 3\,(2\al+1)/4 \ ,
                        \label{T0}
\ea
where $A_{\nu}=2^{\nu}\Gamma(\nu+1)$ and $\Ju(z)$ is a Bessel function
of the first kind. The second linearly independent solution is
ruled out since $Y_{\nu}(k\eta)\to\ift$ as $\k\to 0$.
{}From \rf{Turok.phi} one finds the structure factor
\be
    \aver{\vpk(\eA)\vdot\vpK(\eB)} =
         A_{\nu}^2\:\frac{\curv{\eA\eB}^{3/2}}{\eZ^3/\Delta}
    \frac{J_{\nu}(k\eA)J_{\nu}(k\eB)}{(k\eA)^{\nu}(k\eB)^{\nu}} \ .
                        \label{Sk}
\ee
To obtain the pair correlation function we Fourier transform \rf{Sk}.
Using the normalization $\ci(\bb{0},\eta,\eta)=1$, we have
\ba
\cire
&\eq& \aver{ \vp(\bb{0},\eA) \vdot \vp(\r,\eB) } \nn \\
& = & \intk\,\aver{\vpk(\eA)\vdot\vpK(\eB)}\,\exprk /
      \intk\,\aver{\vpk(\eA)\vdot\vpK(\eA)} \nn \\
& = & \frac{\curv{\eA\eB}^{3/2}}{N_1}\:\intk\,
      \frac{\Ju(k\eA)\Ju(k\eB)}{(k\eA)^{\nu}(k\eB)^{\nu}}\exprk
                        \label{FT}
\ea
\be
      N_1 \eq \inty\,\curv{\frac{\Ju(y)}{y^{\nu}}}^2
            = \frac{\pi\:B\curv{\al,3/2}}
                   {2^{\al\!-\!2}\:\Gamma\curv{\frac{\al+1}{2}}^2} \ .
                        \label{N1}
\ee
Clearly, from \rf{FT}, $\cire$ has the scaling form $\fin(x,q)$,
with $x=r/\eA$ and $q=\eB/\eA$.
To evaluate \rf{FT} we write it in the convenient form
\ba
\fin(x,q)
& = & \frac{q^{3/2}}{N_1}\:
      \inty\,\frac{\Ju(y)\Ju(yq)}{y^{\nu}(yq)^{\nu}}\expxy \nn \\
& = & \frac{2\pi\,q^{3/2}}{N_1}\:\curv{-\frac{1}{x}\,\frac{d}{dx}}\:
      \intinfy\,\cos(yx)\frac{\Ju(y)\Ju(yq)}{y^{\nu}(yq)^{\nu}} \ .
\ea
Using the integral representation of Bessel functions \cite{AS}, we
have
\ba
\!\!\!\!\!\!\!\!\!\!\!\!\!\!\!\!\!\!\!\!
I(x,q) &\eq&
  -\frac{2}{\pi(1\!+\!\al)}\,2^{2\nu-1}\pi\Gamma\curv{\frac{\nu+1/2}{2}}
      \,\intinfy\,\cos(yx)\frac{\Ju(y)\Ju(yq)}{y^{\nu}(yq)^{\nu}} \nn \\
       & = & -\frac{2}{\pi(1\!+\!\al)}\:\intinfy\int^1_0\!ds\int^1_0\!dt\,
      (1-s^2)^{\nu-1/2}(1-t^2)^{\nu-1/2}\cos(yx)\cos(ys)\cos(yqt) \nn \\
       & = & -\frac{1}{(1\!+\!\al)\,q^{\al+2}}
      \int^1_0\!ds\int^q_0\!dt\,(1\!-\!s^2)^{\aP}(q^2\!-\!t^2)^{\aP}
\brak{\dlt(x\pp s\mm t)+\dlt(x\mm s\pp t)+\dlt(\mm x\pp s\pp t)} \nn \\
       & = & -\frac{1}{(1\!+\!\al)\,q^{\al+2}}\:
              \int^1_{-1}\!ds\,(1-s^2)^{\aP}(q^2-(x-s)^2)^{\aP}
              \theta(s+q-x)\theta(q+x-s)
                        \label{Ia} \\
       & = & -\frac{\theta(1\pp q\mm x)}{(1\!+\!\al)\,q^{\al+2}}\:
              \int^B_A\!ds\,(1-s^2)^{\aP}(q^2-(x-s)^2)^{\aP} \ ,
                        \label{Iaa}
\ea
alternatively, performing the `self-similar' transform
$s\leftrightarrow x-s$ in \rf{Ia},
\ba
I(x,q) & = & -\frac{1}{(1\!+\!\al)\,q^{\al+2}}\:
              \int^{x+1}_{x-1}\!ds\,(1-(x-s)^2)^{\aP}(q^2-s^2)^{\aP}
              \theta(q-s)\theta(q+s) \nn \\
       & = & -\frac{\theta(1\pp q\mm x)}{(1\!+\!\al)\,q^{\al+2}}\:
              \int^{x-A}_{x-B}\!ds\,(1-(x-s)^2)^{\aP}(q^2-s^2)^{\aP} \ ,
                        \label{Ibb}
\ea
\ba
    (B,A) & = & (x+q,x-q)   \ \ , \ \ x \leq 1-q \nn \\
          & = & (1,  x-q)   \ \ , \ \ |1-q| \leq x \leq 1+q \nn \\
          & = & (1,   -1)   \ \ , \ \ x \leq q-1 \ ,
                        \label{ABx}
\ea
Differentiating $I(x,q)$ with respect to $x$ we get, using some of the
possible integral representations for $\partial I/\partial x$,
\ba
\!\!\!\!\!\!\!\!\!\!\!\!\!\!\!
\fin(x,q) \! & \! = \! &
       \frac{q^{3/2}}{N\:x}\:\frac{\partial I(x,q)}{\partial x}
                        \label{fdI} \nn \\
               & \! =\! &
       \frac{\theta(1\pp q\mm x)}{N\:x\,q^{\al+1/2}}\:
              \int^B_A\!ds\,(x-s)(q^2-(x-s)^2)^{\aM}(1-s^2)^{\aP}
                        \label{fa} \\
               & \! =\! &
       \frac{\theta(1\pp q\mm x)}{N\:x\,q^{\al+1/2}}\:
              \int^B_A\!ds\,s\,(1-s^2)^{\aM}(q^2-(x-s)^2)^{\aP}
                        \label{fb} \\
               & \! =\! &
      \frac{\theta(1\pp q\mm x)}{N\:x\,q^{\al+1/2}}\,
\int^B_A\!ds\,\frac{\brak{s(q^2\mm 1)\pp x\curv{1\mm s(x\mm s)}}}{2}
                           (1\mm s^2)^{\aM}(q^2\mm (x\mm s)^2)^{\aM} \ ,
                        \label{fc}
\ea
where $N=(\al+1)\,B\curv{\al,3/2}$.
Expression \rf{fa} follows from differentiating \rf{Iaa}.
The form \rf{fb}, which follows from \rf{Ibb} and the transformation
$s\to x-s$, or from integrating \rf{fa} by parts,
is convenient for further differentiation with respect to $x$.
Finally, \rf{fc} is the mean of the previous two, and proves useful at
equal-times ($q=1$) where the factor $1/x$ gets canceled and higher
derivatives with respect to $x$ become easier to evaluate.

By construction $I(x,q)$ must be invariant under interchange of times,
i.e. $I(x,q)=I(x/q,1/q)$. Since it is not explicitly symmetric,
a number of integration variable changes and other transformations may
be performed in $I(x,q)$ and $dI(x,q)/dx$ leading to different
equivalent integral representations for $\fin$.
However, the expressions given, with three different integration limits
\rf{ABx} depending on $x$ and $q$, admit no further simplification.
Writing the integrand in, say, \rf{Iaa}, as $\zeta(x,q;s)^{\aP}$, it is
easy to see that $\zeta(x,q;s)=(1-s)(1+s)(q+x-s)(q-x+s)$ is
{\em non-negative} and {\em bounded} only in the regions where both
$|s|\leq 1$ and $x-q\leq s\leq x+q$, which are precisely those yield by
\rf{ABx}. Hence, since $\aP$ is non-integer, the integral \rf{Iaa} runs
over the whole (bounded) region where the integrand is {\em real}.
As illustrated by the small-$x$ expansion \rf{small-x.fi}, $I(x,q)$ is
singular for $\al=2$. In fact, at each integration limit one (or
two, if $x=0$) of the radicals in $\zeta(x,q;s)$ vanishes and high
enough derivatives of the integrand or integration limits will diverge.
Up to fourth order, however, we get finite derivatives of $I(x,q)$, but
since each of the radicals in \rf{Iaa} can only be differentiated twice,
one has to transform the integral, e.g. using $r-s\to s$ (or integrating
by parts) before doing the third and fourth derivatives. Using these
methods we find, form \rf{fi}, with $\delta\eq\al+1/2$,
\ba
 \ci(\dot{1},2) & = &       \frac{1}{\eA}\,\curv{
  \frac{1}{N\:x\,q^{\delta}}\:\int^B_A\!ds\,F_{x,1}-\delta\,\ci(1,2) }
                        \label{DtA} \nn \\
 \ci(1,\dot{2}) & = &       \frac{1}{\eB}\,\curv{
  \frac{1}{N\:x\,q^{\delta}}\:\int^B_A\!ds\,q\,F_{x,2}-\delta\,\ci(1,2) }
                        \label{DtB} \nn \\
\ci(\dot{1},\dot{2}) & = &
\!\!\!\!\frac{1}{\eA\eB}\curv{
\frac{1}{N\:x\,q^{\delta}}\int^B_A\!ds\,q\,F_{x,1,2}\!-\!\delta^2\ci(1,2)
\!-\!\delta\curv{ \eta_1\ci(\dot{1},2)\!+\!\eta_2\ci(1,\dot{2}) } }
                        \label{DtADtB} \nn \\
\!\!\!\!\!\!\!\!\!\!\!\!\!
 \frac{\ptl\ci(1,2)}{\ptl r} & = &  \frac{1}{\eA}\,\curv{
 \frac{1}{N\:x\,q^{\delta}}\:\int^B_A\!ds\,F_{x^2}-\frac{\ci(1,2)}{x} }
                        \label{DelDel} \nn \\
\nb^2\ci(1,2) & = & \frac{1}{\eA^2}\,\frac{1}{N\:x\,q^{\delta}}\:
              \int^B_A\!ds\,F_{x^3}
                        \label{DelB}
\ea
where it is implicit that $x\leq 1+q$, and
\ba
   F_{x^2}   & = &
(q^2-\al(x-s)^2)(1-s^2)^{(\al+1)/2}(q^2-(x-s)^2)^{(\al-3)/2} \nn \\
   F_{x^3}   & = & -(\al+1)\,
s\,(q^2-\al(x-s)^2)(1-s^2)^{(\al-1)/2}(q^2-(x-s)^2)^{(\al-3)/2} \nn \\
   F_{x,1}   & = & (\al+1)\,
(x-s)(1-s^2)^{(\al-1)/2}(q^2-(x-s)^2)^{(\al-1)/2} \nn \\
   F_{x,2}   & = & (\al-1)\,q\,
(x-s)(1-s^2)^{(\al+1)/2}(q^2-(x-s)^2)^{(\al-3)/2} \nn \\
   F_{x,1,2} & = & (\al^2-1)\,q\,
(x-s)(1-s^2)^{(\al-1)/2}(q^2-(x-s)^2)^{(\al-3)/2} \ .
                        \label{integrands}
\ea
We also obtain in the limit $2\to 1$, i.e. $\r\to 0$ and $\eB\to\eA$,
\ba
 \ci(\dot{1},1) = 0 & , & \nb\ci(1,1) = 0 \nn \\
 \ci(\dot{1},\dot{1}) & = & \frac{1}{\eA^2}\,\frac{T_0}{\al-2}
                        = - \nb^2\ci(1,1) + \frac{T_0}{\eta_1^2} \nn \\
 - \nb^2\ci(1,1) & = & \frac{\al-1}{\eA^2}\,\frac{T_0}{\al-2} \ .
                        \label{limits}
\ea

\newpage

\singlespacing



\begin{thebibliography}{99}
\bibitem{KYGNote} This approach is mathematically equivalent to the
approximations used in references \cite{OJK,KYG,BPT} for condensed
matter systems described by `Model A' dynamics.
\end{thebibliography}
\end{document}